\documentclass[letterpaper, 10 pt,conference]{cssconf}  

\IEEEoverridecommandlockouts
\usepackage{amssymb}
\usepackage{amsmath}
\usepackage{graphicx,color}
\usepackage{enumerate}
\usepackage{lipsum} 
\usepackage[capitalise]{cleveref}
\usepackage{tikz}
\usetikzlibrary{patterns}
\usepackage{accents}

\usepackage{enumerate}
\usepackage{comment}
\usepackage{bm}
\usepackage{algorithm}
\usepackage{algorithmic}

\usepackage{url}

\usepackage{cite}

\usepackage{soul}

\newtheorem{proposition}{Proposition}
\newtheorem{remark}{Remark}
\newtheorem{theorem}{Theorem}
\newtheorem{definition}{Definition}

\newtheorem{lemma}{Lemma}
\newtheorem{example}{Example}

\renewcommand{\baselinestretch}{0.94}

\DeclareMathOperator{\Equaldef}{\overset{def}{=}}

\renewcommand{\Pr}{\mathbf{P}}

\title{\LARGE \bf Robust remote estimation over the collision channel \\
in the presence of an intelligent jammer}
\author{Xu Zhang and 
Marcos M. Vasconcelos
\thanks{X. Zhang is with LSEC, Academy of Mathematics and Systems Science, Chinese Academy of Sciences, Beijing, China. X. Zhang was supported by China National Postdoctoral Program
	for Innovative Talents under Grant No. BX2021346. E-mail: \texttt{xuzhang\_cas@lsec.cc.ac.cn}.}
\thanks{M. M. Vasconcelos is with the Commonwealth Cyber Initiative and the Bradley Department of Electrical and Computer Engineering, Virginia Tech.
M. M. Vasconcelos was supported by funds from the Commonwealth Cyber Initiative (CCI).
E-mail: \texttt{marcosv@vt.edu}.}
}

\begin{document}

\maketitle

\begin{abstract}

We consider a sensor-receiver pair communicating over a wireless channel in the presence of a jammer who may launch a denial-of-service attack. We formulate a zero-sum game between a coordinator that jointly designs the transmission and estimation policies, and the jammer. We consider two cases depending on whether the jammer can sense the channel or not. We characterize a saddle-point equilibrium for the class of symmetric and unimodal probability density functions when the jammer cannot sense the channel. If the jammer can sense if the channel is being used, we provide an efficient algorithm that alternates between iterations of Projected Gradient Ascent and the Convex-Concave Procedure to find approximate First-order Nash-Equilibria. Our numerical results show that in certain cases the jammer may decide to launch a denial-of-service attack with the goal of deceiving the receiver even when the sensor decides not to transmit.  

\end{abstract}

\section{Introduction}

Cyber-Physical Systems are characterized by the tight coupling between physical, computing and communication components. Due to the confluence of three distinct branches of systems science, many new security vulnerabilities have emerged over the last decade as cyber-physical systems started to dominate the technology landscape \cite{Pasqualetti:2015}. Many of the critical infrastructures our society relies on are cyber-physical systems (e.g. industrial automation systems, transportation networks, utility distribution networks, etc.). In particular, remote sensing where one (or multiple) sensor(s) communicates its measurements over a wireless channel to a non-collocated access point or base-station is a fundamental building block of many cyber-physical systems \cite{Vasconcelos:2020}. The openness of the wireless medium creates a vulnerability to attacks that could compromise the performance and safe operation of the entire system \cite{griffioen2019tutorial}.  

Denial-of-Service (DoS) is a class of cyber-attacks where a malicious agent, often referred to as the \textit{jammer}, may disrupt the communication link between the legitimate transmitter-receiver pair. DoS attacks are widely studied at different levels of modeling detail of the communication channel. For example if the channel is assumed to be a physical layer model, the jammer may introduce additional Gaussian noise to the transmitted signal. If the channel is modeled at the network layer by a packet-drop channel, the jammer may increase the probability of dropping a packet. We consider a medium access control (MAC) layer model in which the jammer may decide to block the channel by transmitting an interference signal that overwhelms the receiver, causing a packet collision. 

\begin{figure}[t!]
    \centering
    \includegraphics[width=0.9\columnwidth]{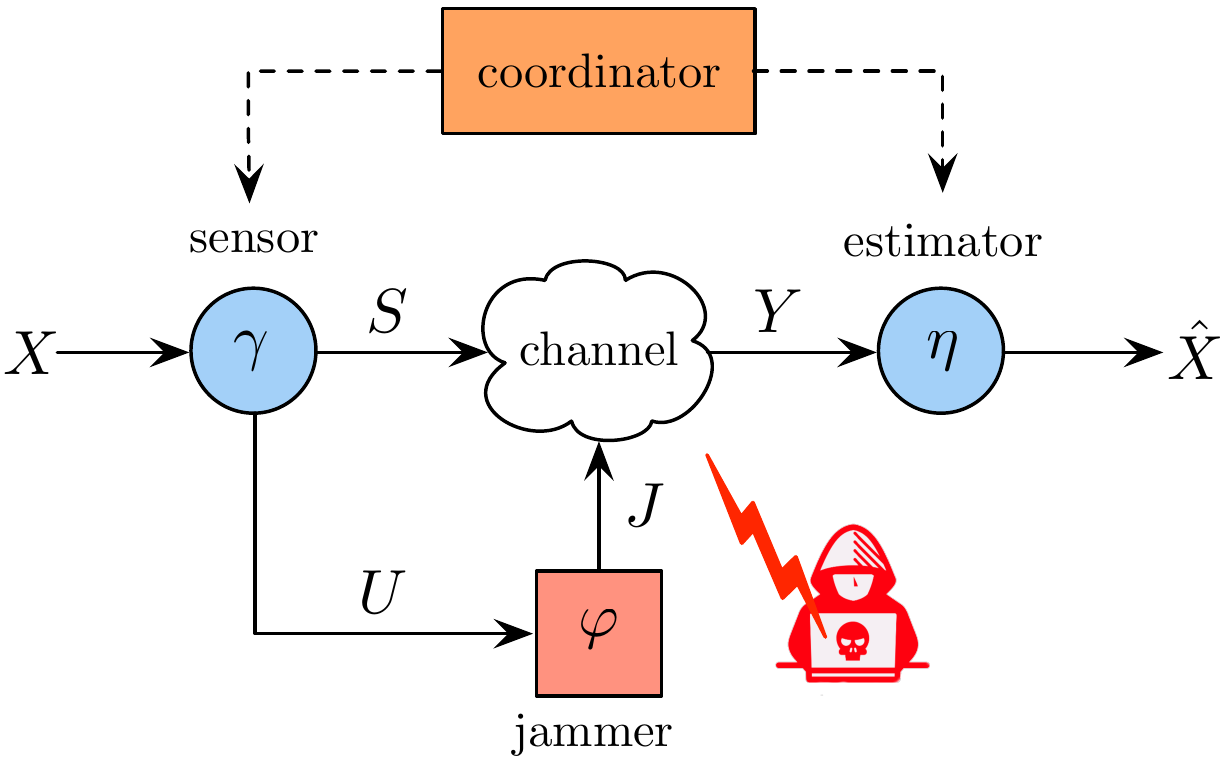}
    \caption{Block diagram for a game between a coordinator and a jammer. The jammer may have access to the sensor's decision to transmit. The coordinator designs the policies for the sensor and the estimator.}
    \label{fig:system}
\end{figure}

We consider the remote estimation system depicted in \cref{fig:system}, which is comprised of a sensor and estimator pair communicating over a collision channel in the presence of a jammer.
The sensor makes a stochastic measurement $X$ of a physical quantity according to a given distribution, and decides whether to transmit it or not to the estimator. Communication is costly, therefore, the sensor must transmit wisely. We consider two cases: 1. the jammer cannot sense if the channel is being used by the sensor; 2. the jammer can sense the channel, i.e., has access to $U$. 
Jamming is assumed to be costly, therefore, the jammer must act strategically. 

Finally, the estimator observes the channel output and declares an estimate $\hat{X}$ for the sensor's observation such as to minimize the expected quadratic distortion between $X$ and $\hat{X}$. We study this problem as a zero-sum game between a coordinator (system designer) and the jammer. Our goal is to characterize equilibrium solutions and obtain efficient algorithms to compute them. The main difference between our model and existing work in this area is the presence of a virtual binary signaling channel that can be exploited by the coordinator to guarantee a minimum level of performance of the system  in the presence of DoS attacks. 

\subsection{Related literature}

There exists an extensive literature on strategic communication in the presence of jammers. This class of problems seems to have started with the work of Basar \cite{basar1983gaussian}, which obtained a complete characterization of the saddle-point equilibria when the sensor measurements and the channel are Gaussian. Recently, an extension to the two-way additive Gaussian noise channel was studied by McDonald et al. in \cite{mcdonald2019two}. A jamming problem where the transmitter and estimator have different objectives was solved by Akyol et al. in \cite{Akyol2017Info} using a hierarchical game approach. A jamming problem with and without common randomness between the transmitter and estimator is studied Akyol in \cite{Akyol2019Optimal} and a Stackelberg game formulation was considered by Gao et al. \cite{Gao2019Communication}. Another interesting problem formulation is due to Shafiee and Ulukus in \cite{Shafiee2009Mutual}, where the pay-off function is the mutual information between the channel input and output. Jamming over fading channels was considered by Ray et al. in \cite{ray2006optimal} and subsequently by Altman et al. in \cite{Altman:2011}. An LTE network model was considered by Aziz et al. in \cite{aziz2020resilience}.

Another class of remote estimation problems focuses on the state estimation of a linear time invariant system driven by Gaussian noise under DoS attacks. Li et al. \cite{li2015jamming} studied a jamming game where the transmitter and jammer have binary actions. A SINR-based model was considered by Li et al. in \cite{li2016sinr}, where the transmitter and jammer decide among multiple discrete power levels. The case of continuum of power levels was studied by Ding et al. in \cite{ding2017stochastic}. \textcolor{black}{A jamming model over a channel with two modes (i.e., free mode and safe mode) was analyzed by Wu et al. in \cite{Wu:2017}.}
A jamming problem with asymmetric feedback information and multi-channel transmissions was considered by Ding et al. in \cite{ding2018attacks} and \cite{ding2017multi}, respectively. A Stackelberg equilibrium approach to this problem was considered by Feng et al. in \cite{Feng:2021}. The problem of optimizing the attack scheduling policy from the jammer's perspective was considered by Peng et al. in \cite{peng2017optimal}.  

The model described herein is closely related to the work of Gupta et al. \cite{gupta2012dynamic,gupta2016} and Vasconcelos and Martins \cite{vasconcelos:2017a,vasconcelos2018optimal}, where there is a clear distinction between the channel being blocked vs. idle. As in \cite{gupta2012dynamic}, we assume that
the transmission decision $U$ may be available to the jammer, but not the full input signal $X$. This assumption is realistic in the sense that the bits used to encode $X$ may be encrypted. In the game considered in \cite{gupta2012dynamic}, it is assumed that the receiver is fixed, and the game is played between the sensor and the jammer. Instead, we follow Akyol \cite{Akyol2019Optimal} in which the sensor and estimator are distinct agents implementing policies optimized by a \textit{coordinator} \cite{Nayyar2014}.

\subsection{Contributions}

The main contributions of this article are:

\begin{itemize}

\item We introduce a new class of signaling games among a sensor, an estimator and a jammer with asymmetric information, over a collision channel model.

\item For a jammer without channel sensing capability, we obtain a sufficient condition on the \textcolor{black}{probability density function} that guarantees the existence of a saddle-point equilibrium despite the overall lack of convexity for the coordinator's sub-problem.

\item For a reactive jammer, we obtain an algorithm based on an interleaved projected gradient ascent and a convex-concave procedure to efficiently obtain an approximate first-order Nash-equilibrium.

\end{itemize}

\section{System model}

We consider the system in \cref{fig:system}, which consists of a non-colocated sensor-estimator pair communicating over a wireless network vulnerable to DoS attacks. 
The sensor observes realizations of a random variable $X$, which is distributed according to a continuous probability density function $f$. We assume that $f$ is supported on the real line, i.e., $f(x)>0$, $x \in \mathbb{R}$. However, the results herein can easily be extended to random vectors. Upon observing $X=x$, the sensor decides whether to transmit its observation or not according to a mixed policy $\gamma:\mathbb{R}\rightarrow [0,1]$, such that
\begin{equation}
\Pr(U=1\mid X=x) = \gamma(x),
\end{equation} 
where the decision variable $U=1$ if the sensor transmits, and $U=0$ if the sensor remains silent. Then, the channel input signal $S$ is determined as
\begin{equation}
S = \begin{cases}
X, & \text{if} \ \ U=1 \\
\varnothing, & \text{if} \ \ U=0,
\end{cases}
\end{equation} 
where $\varnothing$ is used to denote that the channel is idle.

The jammer operates strategically based on side information about the the sensor's decision to transmit or not. Upon observing $U=u$, the jammer blocks the channel using a mixed policy $\varphi:\{0,1\}\rightarrow [0,1]$ such that
\begin{equation}
\Pr(J=1 \mid U=u) = \varphi(u),
\end{equation}
where $J=1$ denotes the jammer's decision to block the channel, and $J=0$ denotes the decision not to block. Unlike \cite{gupta2012dynamic}, we are not precluding the jammer to block an idle channel.

Given the input signal $S$ and the jammer's decision $J$, the channel ouput $Y$ is given by
\begin{equation} \label{eq: Y}
Y =\begin{cases}
S & \text{if} \ \ J=0 \\
\mathsf{B} & \text{if} \ \ J=1,
\end{cases}
\end{equation}
where $\mathsf{B}$ denotes that the channel has been blocked, and a DoS attack has occurred.
	
Finally, the receiver uses an estimation policy $\eta: \mathbb{R}\cup\{\varnothing,\mathsf{B}\} \to \mathbb{R}$ such that
\begin{equation} \label{eq: estimator}
\eta(y) =\begin{cases}
x & \text{if} \ \ y=x \\ 
\hat{x}_0 & \text{if}\ \ y=\varnothing\\ 
\hat{x}_1 & \text{if}\ \ y=\mathsf{B},
\end{cases}
\end{equation}
where $\hat{x}_0$ and $\hat{x}_1$ are the so-called representation symbols used by the receiver when the channel is idle and blocked, respectively. This is another departure from the model in \cite{gupta2012dynamic}, which does not account for such distinction.  For the remainder of the paper, let $\hat{x} \Equaldef  (\hat{x}_0,\hat{x}_1)$.

We consider the same objective function of \cite{gupta2012dynamic}, which consists of three terms: the estimation error, and the communication and jamming costs, as follows
\begin{equation}\label{eq:objective}
\mathcal{J}\big( (\gamma,\eta),\varphi \big) = \mathbf{E}\big[ (X-\hat{X})^2 \big] + c\Pr(U=1) - d\Pr(J=1). 
\end{equation}

Even though the sensor and the estimator act as independent agents, we adopt a zero-sum game between a coordinator that seeks to minimize \cref{eq:objective} by jointly designing the pair $(\gamma,\eta)$, and the jammer whose goal is to maximize \cref{eq:objective} with respect to $\varphi$.

\section{Jamming without sensing the channel}

We start our analysis by considering the jammer without channel sensing, which does not have access to the side information $U$. Since the jammer does not have access to $U$, it randomly blocks the channel with some fixed probability $\varphi\in[0,1]$, i.e., $
\mathbf{P}(J=1) = \varphi.$ More importantly, the random variable $J$ is independent of $U$, i.e., $J\perp \!\!\! \perp U$. 

We are interested in obtaining policies tuples $(\gamma^\star,\eta^\star,\varphi^\star)$ that constitute a saddle-point equilibrium, i.e.,
\begin{equation}\label{eq:SPE}
\mathcal{J}\big( (\gamma^\star,\eta^\star),\varphi \big) \leq \mathcal{J}\big( (\gamma^\star,\eta^\star),\varphi^\star \big) \leq \mathcal{J}\big( (\gamma,\eta),\varphi^\star \big),
\end{equation}
for all $\gamma,\eta,\varphi$ in their respective admissible policy spaces.

The first step is to obtain a structural result for the set of optimal transmission strategies for the sensor.

\vspace{5pt}

\begin{proposition}[Optimality of threshold policies]\label{prop:threshold} For a system with a jammer with fixed jamming probability $\varphi \in[0,1]$, and an arbitrary estimation policy indexed by representation symbols $\hat{x}\in\mathbb{R}^2$, the optimal transmission strategy is\footnote{The function $\mathbf{1}(\mathfrak{S})$ denotes the indicator function of the Boolean statement $\mathfrak{S}$, i.e., $\mathbf{1}(\mathfrak{S})=1$ if $\mathfrak{S}$ is true, and $\mathbf{1}(\mathfrak{S})=0$ if $\mathfrak{S}$ is false.}:
\begin{equation}
\gamma_{\eta,\varphi}^\star(x) = 
\mathbf{1}\big( (1-\varphi)(x-\hat{x}_0)^2 \geq c \big).
\end{equation}
\end{proposition}

\vspace{5pt}


\vspace{5pt}

\begin{proof} 
Using the law of total expectation and the definition of the estimation policy in \cref{eq: estimator},  we rewrite \cref{eq:objective} as follows:
\begin{multline}
\mathcal{J}\big( (\gamma,\eta),\varphi \big) = \\
\mathbf{E}\big[(X-\hat{x}_0) \mid U=0,J=0 \big]\Pr(U=0,J=0) \\
+ \mathbf{E}\big[(X-\hat{x}_1) \mid U=0,J=1 \big]\Pr(U=0,J=1) \\
+ \mathbf{E}\big[(X-\hat{x}_1) \mid U=1,J=1 \big]\Pr(U=1,J=1) \\
+ c\Pr(U=1) -d\Pr(J=1).  
\end{multline}
From the fact that $(U,X) \perp \!\!\! \perp J$ and $\Pr(J=1)=\varphi$, we have:
\begin{multline}
\mathcal{J}\big( (\gamma,\eta),\varphi \big) = 
\mathbf{E}\big[(X-\hat{x}_0) \mid U=0 \big]\Pr(U=0)(1-\varphi) \\ 
+ \mathbf{E}\big[(X-\hat{x}_1) \big]\varphi + c\Pr(U=1) -d\varphi,  
\end{multline}
which is equivalent to
\begin{multline}
\mathcal{J}\big( (\gamma,\eta),\varphi \big) = \int_\mathbb{R} (1-\varphi)(x-\hat{x}_0)^2\big((1-\gamma(x)\big)f(x)\mathrm{d} x \\
+ \int_{\mathbb{R}}c\gamma(x)f(x)\mathrm{d} x + \varphi \mathbf{E}\big[(X-\hat{x}_1) \big] -d\varphi.
\end{multline}
Finally, when optimizing over $\gamma$ for fixed $\hat{x}$ and $\varphi$, we have an infinite dimensional linear program with the following constraint:
\begin{equation}
0 \leq \gamma(x) \leq 1,  \ \ x\in\mathbb{R}.
\end{equation}
The solution to this problem is obtained by comparing the arguments of the two integrals, i.e., $x \in \{\xi \mid \gamma^\star_{\eta,\varphi}(\xi) = 0\}$ if and only if 
\begin{equation}
(1-\varphi)(x-\hat{x}_0)^2 \leq c.
\end{equation}
\end{proof}

\vspace{5pt}

\begin{remark}
\Cref{prop:threshold} implies that the optimal transmission policy is always of the threshold type. \textcolor{black}{This threshold policy is symmetric only if $\hat{x}_0=0$}. The optimal policy will be characterized by finite lower and upper thresholds, if $\varphi \in[0,1)$ or a degenerate policy called \textit{never transmit} when $\varphi=1$.
\end{remark}

\vspace{5pt}

The structure of the optimal transmission policy in \cref{prop:threshold} implies that the objective function assumes the following expression:
\begin{multline}\label{eq:new_objective}
\mathcal{J}\big((\gamma_{\eta,\varphi}^\star,\eta),\varphi\big) = \mathbf{E}\bigg[\min\Big\{(1-\varphi)(X-\hat{x}_0)^2, c \Big\} \bigg] \\ + \varphi\Big(\mathbf{E}\big[(X-\hat{x}_1)^2 \big] -d \Big).
\end{multline}

\vspace{5pt}

The second step in the analysis is to optimize over the estimation strategy, which is a finite dimensional optimization problem over $\hat{x}\in\mathbb{R}^2$. The pair $(\hat{x}_0^\star,\hat{x}_1^\star)$ that minimizes \cref{eq:new_objective} depends on the pdf $f$, the constants $c,d$ and the strategy of the jammer. It is easy to see that $\hat{x}_1^\star = \mathbf{E}[X]$. However, the optimal value of $\hat{x}_0^\star$ does not admit a closed form expression and must be found numerically, in general. However, in the following result we establish a condition on the pdf $f$ such that the optimal $\hat{x}_0^\star=\mathbf{E}[X]$ for any $c, d$ and $\varphi$.  

\vspace{5pt}

\begin{theorem} \label{thm: optimal_estimator}
If $f(x)$ is a symmetric and unimodal pdf around $\mathbf{E}[X]$, then 
\begin{equation}
\eta^\star (y) = \begin{cases}
\mathbf{E}[X], & \text{if} \ \ y \in \{\varnothing,\mathsf{B}\} \\
x, & \text{if} \ \ y = x.
\end{cases}
\end{equation}
\end{theorem}

\vspace{5pt}

The proof of this result requires the following definitions.

\vspace{5pt}

{ \color{black}
\begin{definition}[Symmetric rearrangement] Let $\mathbb{D} \subset \mathbb{R}$ be  a set of finite measure. Its symmetric rearrangement $\mathbb{D}^*$ is defined as the open interval centered at the origin whose measure is the same as $\mathbb{D}$.
\end{definition}
\vspace{5pt}
\begin{definition}[Symmetric decreasing rearrangement] Let $f:\mathbb{R} \to \mathbb{R}$ be a nonnegative measurable function that vanishes at infinity. Its symmetric decreasing rearrangement $f^\downarrow$ is
\begin{equation}
	f^{\downarrow}(x) \Equaldef \int_{0}^{\infty} \mathbf{1}\left(x \in\left\{\xi \in \mathbb{R} \mid f(\xi)>t\right\}^{*}\right) \mathrm{d} t.
\end{equation}
	
\end{definition}
}
\vspace{5pt}

\begin{lemma}[Hardy-Littlewood Inequality \cite{Burchard:2009}] Let $f: \mathbb{R} \to \mathbb{R} $ and $g: \mathbb{R} \to \mathbb{R}$ be nonnegative measurable functions that vanish at infinity. The following inequality holds: 
		\begin{equation}
			\int_{\mathbb{R}} f(x)g(x) \mathrm{d} x \le \int_{\mathbb{R}} f^{\downarrow}(x)g^{\downarrow}(x) \mathrm{d} x,
		\end{equation}
where $f^\downarrow$ and $g^\downarrow$ are the symmetric decreasing rearrangements of $f$ and $g$, respectively. 

\end{lemma}

\vspace{5pt}

\begin{proof}\textit{(Proof of Theorem 1)}
From \cref{eq:new_objective}, it is easy to show that $\hat{x}_1^\star = \mathbf{E}[X]$. Without loss of generality, we assume that $\mathbf{E}[X]=0$. We focus on fixing $\varphi\in[0,1)$ and solving the (equivalent) non-convex optimization problem
\begin{equation}
\min_{\hat{x}_0\in \mathbb{R}} \ \ 	 \int_{\mathbb{R}}\min\Big\{(x-\hat{x}_0)^2, \frac{c}{1-\varphi} \Big\}f(x)\mathrm{d}x.
\end{equation}



Our proof hinges on establishing the following inequality, under the symmetry and unimodality assumption of $f$, 
\begin{multline} \label{neq: h_x_0}
	 \int_{\mathbb{R}}\min\Big\{(x-\hat{x}_0)^2, \frac{c}{1-\varphi} \Big\}f(x)\mathrm{d} x    \\  \geq \int_{\mathbb{R}}\min\Big\{x^2, \frac{c}{1-\varphi} \Big\}f(x)\mathrm{d} x, \ \ \hat{x}_0 \in \mathbb{R}.
\end{multline}
Therefore, implying that $\hat{x}_0^\star=0$. Next, we prove \cref{neq: h_x_0} by considering the following equivalent inequality
\begin{multline}
	\frac{c}{1-\varphi}-\int_{\mathbb{R}}\min\Big\{(x-\hat{x}_0)^2, \frac{c}{1-\varphi} \Big\}f(x)\mathrm{d} x \\ \le  \frac{c}{1-\varphi}-\int_{\mathbb{R}}\min\Big\{x^2, \frac{c}{1-\varphi} \Big\}f(x)\mathrm{d} x,
\end{multline}
which can be represented by
\begin{multline}\label{eq:ineq1}
	 \int_{\mathbb{R}}\max\Big\{\frac{c}{1-\varphi}-(x-\hat{x}_0)^2, 0 \Big\}f(x)\mathrm{d} x \\ \le  
	 \int_{\mathbb{R}}\max\Big\{\frac{c}{1-\varphi}-x^2, 0 \Big\}f(x)\mathrm{d} x. 
\end{multline}
Define
\begin{equation}
g(x;\hat{x}_0) \Equaldef \max\Big\{{c}/(1-\varphi)-(x-\hat{x}_0)^2, 0 \Big\}
\end{equation}
and notice that $g^{\downarrow}(x;\hat{x}_0) = g(x;0)$\footnote{\color{black} When a function is symmetric and unimodal around a non-zero point (in this case $\hat{x}_0$), its symmetric decreasing rearrangement corresponds to shifting the function to \textcolor{black}{the origin.}}. Moreover, from the assumption that $f$ is symmetric and unimodal function implies that $f^{\downarrow}(x) = f(x)$. Since $g(x;\hat{x}_0)$ and $f(x)$ are nonnegative and vanish at infinity, we may use the Hardy-Littlewood inequality, which implies in \cref{eq:ineq1} and, equivalently, in \cref{neq: h_x_0}.

%

\end{proof}

The symmetry and unimodality assumptions on $f$ lead to closed form characterizations for the optimal strategy of the coordinator. These assumptions are common in the remote estimation literature (e.g. \cite{gupta2012dynamic,vasconcelos2017optimal,vasconcelos:2017a}, and references therein), and encompass a large class of distributions, including Gaussian and Laplace. Without loss of generality, for the remainder of the paper we assume that $\mathbf{E}[X]=0$. 

The optimal transmitter and estimator's strategies for a symmetric and unimodal density, implies that the objective function for the jammer is given by
	\begin{multline}\label{eq:jammer_obj}
		\mathcal{J}\big((\gamma_{\eta^\star,\varphi}^\star,\eta^\star),\varphi\big) = \mathbf{E}\bigg[\min\Big\{(1-\varphi) X^2, c \Big\} \bigg] \\ + \varphi\Big(\mathbf{E}\big[X^2 \big] -d \Big).
	\end{multline}
The objective function in \cref{eq:jammer_obj} is concave with respect to $\varphi$, and 
we can explicitly determine the optimal jamming probability $\varphi$ by calculating its derivative and setting it to $0$.

\vspace{5pt}

\begin{theorem} \label{thm: optimal_jam_prob}
	If $f(x)$ is a symmetric and unimodal 
	pdf such that $\mathbf{E}[X]=0$. The optimal jamming probability \textcolor{black}{ under} the optimal transmission policy in \cref{prop:threshold} and  the optimal estimation policy in \cref{thm: optimal_estimator} is
	\begin{equation}
		\varphi^\star=
		\left\{\begin{array}{ll}
			{0} & {\text { if }  2\int_{\sqrt{c}}^{+\infty} x^2f(x)\mathrm{d} x <d} 
			\\ {\tilde{\varphi}} & {\text { if }  2\int_{\sqrt{c}}^{+\infty} x^2f(x)\mathrm{d} x \ge d}
		\end{array}\right.,
	\end{equation}
where $\tilde{\varphi}$ is the unique constant in $[0,1)$ that satisfies
\begin{equation} \label{eq: J_d_beta_tilde}
	 2\int_{\sqrt{c/(1-\tilde{\varphi})}}^{+\infty} x^2f(x)\mathrm{d} x =d. 
\end{equation}
\end{theorem}
\vspace{5pt}
\begin{proof}
	First, we represent \cref{eq:jammer_obj} in integral form as
	\begin{multline}
		\mathcal{J}\big((\gamma_{\eta^\star,\varphi}^\star,\eta^\star),\varphi\big) = \int_{-\sqrt{c/(1-\varphi)}}^{\sqrt{c/(1-\varphi)}} (1-\varphi)x^2f(x)\mathrm{d} x \\ + 2 \int_{\sqrt{c/(1-\varphi)}}^{+\infty} c f(x) \mathrm{d}x 
		+\varphi \Big(\mathbf{E}\big[X^2 \big] -d \Big).
	\end{multline}
		Taking the derivative of  the objective function with respect to $\varphi$, we have
	\begin{IEEEeqnarray}{rCl} \label{eq: J_d_beta}
		\mathcal{G}(\varphi) & \Equaldef & \frac{\partial }{\partial \varphi}\mathcal{J}\big((\gamma_{\eta^\star,\varphi}^\star,\eta^\star),\varphi\big)  \\
		& = & 2\int_{\sqrt{c/(1-\varphi)}}^{+\infty} x^2f(x)\mathrm{d} x -d.
	\end{IEEEeqnarray}
	
	Notice that $\mathcal{G}(\varphi)$ is a monotonically decreasing function with respect to $\varphi$ and
	\begin{align}
		\mathcal{G}(0) & =   2\int_{\sqrt{c}}^{+\infty} x^2f(x)\mathrm{d} x -d. \\
		\lim_{\varphi\uparrow 1}\mathcal{G}(\varphi) & =   -d.
	\end{align}
	
	If $\mathcal{G}(0)\ge0$, then the optimal $\varphi^\star = \tilde{\varphi}$ due to the fact that $\mathcal{G}(\tilde{\varphi})= 0$.	
	If $\mathcal{G}(0)<0$, the objective function decreases with the increasing of $\varphi$. Therefore, $\varphi^\star=0$.
\end{proof}

\Cref{thm:saddle-point} summarizes the saddle-point strategy for the game between a coordinator jointly designing the transmission and estimation strategy against the jammer.

\vspace{5pt}

\begin{theorem}[Saddle-point equilibria]\label{thm:saddle-point} Given a symmetric and unimodal  pdf $f$ with $\mathbf{E}[X]=0$, communication and jamming costs $c,d\geq 0$, the saddle-point strategy $(\gamma^\star,\eta^\star,\varphi^\star)$ for the game with jammer without channel sensing is given by:	
	\begin{itemize} 
		\item[1)] If $2\int_{\sqrt{c}}^{+\infty} x^2f(x)\mathrm{d} x <d$, the optimal policies are 
		\begin{align}\label{eq:saddle_pt_11}
			&\gamma^\star(x) = \mathbf{1}(x^2 > c)\\ 
			&\eta^\star (y) = \begin{cases}
				0, & \text{if} \ \ y \in \{\varnothing,\mathsf{B}\} \\
				x, & \text{if} \ \ y = x, 
			\end{cases} \label{eq:saddle_pt_12} \\
			&\varphi^\star =0.
		\end{align}
		
		\item[2)] If $2\int_{\sqrt{c}}^{+\infty} x^2f(x)\mathrm{d} x \ge d$, the optimal policies are
		\begin{align}\label{eq:saddle_pt_21}
			&\gamma^\star(x) = \mathbf{1}\big((1-\tilde{\varphi})x^2 > c\big) \\ 
			&\eta^\star (y) = \begin{cases}
				0, & \text{if} \ \ y \in \{\varnothing,\mathsf{B}\} \\
				x, & \text{if} \ \ y = x 
			\end{cases} \label{eq:saddle_pt_22} \\
			&\varphi^\star=\tilde{\varphi},
		\end{align}
		where $\tilde{\varphi}$ is the unique solution of \cref{eq: J_d_beta_tilde}.
	\end{itemize}
\end{theorem}

\vspace{5pt}

\begin{proof} We consider two cases:
	
		Case 1 -- Assume that $2\int_{\sqrt{c}}^{+\infty} x^2f(x)\mathrm{d} x <d$. If the jammer chooses not to block the channel, i.e., $\varphi^\star=0$, using \cref{prop:threshold} we have the corresponding optimal transmission strategy
		\begin{equation}
				\gamma^\star(x) = \mathbf{1}(x^2 > c). 
		\end{equation}
		Under the jammer and transmitter's policies above, \cref{thm: optimal_estimator} yields that $\hat{x}^\star_0=0$ and $\hat{x}_1^\star=0$. In this case, the policies satisfy
		 $
		 	\mathcal{J}\big( (\gamma^\star,\eta^\star),\varphi^\star \big) \le \mathcal{J}\big( (\gamma,\eta),\varphi^\star \big).
		 $
		 
	 	If the optimal transmission strategy is $\gamma^\star$ and the optimal estimator is $\eta^\star$, using \cref{thm: optimal_jam_prob} and the assumption that $2\int_{\sqrt{c}}^{+\infty} x^2f(x)\mathrm{d} x <d$, we get the optimal jammer's strategy is $\varphi^\star=0$. Therefore, we have
	 	 $
	 	\mathcal{J}\big( (\gamma^\star,\eta^\star),\varphi \big) \le \mathcal{J}\big( (\gamma^\star,\eta^\star),\varphi^\star \big).
	    $

		Case 2 -- Assume that $2\int_{\sqrt{c}}^{+\infty} x^2f(x)\mathrm{d} x \ge d$. If the jammer blocks the channel with probability $\tilde{\varphi}$, using \cref{prop:threshold} gives the corresponding optimal transmission strategy
		\begin{equation}
			\gamma^\star(x) = \mathbf{1}\big((1-\tilde{\varphi})x^2 > c\big).
		\end{equation}
		Under the jammer and transmitter's policies above, using \cref{thm: optimal_estimator} yields that $\hat{x}_0=0$ and $\hat{x}_1=0$. Therefore, we have
	    $
			\mathcal{J}\big( (\gamma^\star,\eta^\star),\varphi^\star \big) \le \mathcal{J}\big( (\gamma,\eta),\varphi^\star \big).
	    $
		
	   If the optimal transmission strategy is $\gamma^\star$ and the optimal estimator is $\eta^\star$, using \cref{thm: optimal_jam_prob} and the assumption that $2\int_{\sqrt{c}}^{+\infty} x^2f(x)\mathrm{d} x \ge d$, we get the optimal jammer's strategy is $\varphi^\star=\tilde{\varphi}$. Therefore, we have
	   $
	   	\mathcal{J}\big( (\gamma^\star,\eta^\star),\varphi \big) \le \mathcal{J}\big( (\gamma^\star,\eta^\star),\varphi^\star \big).
	   $
	
\end{proof}

\begin{figure}[!t]
	\centering
	\includegraphics[width=0.9\columnwidth]{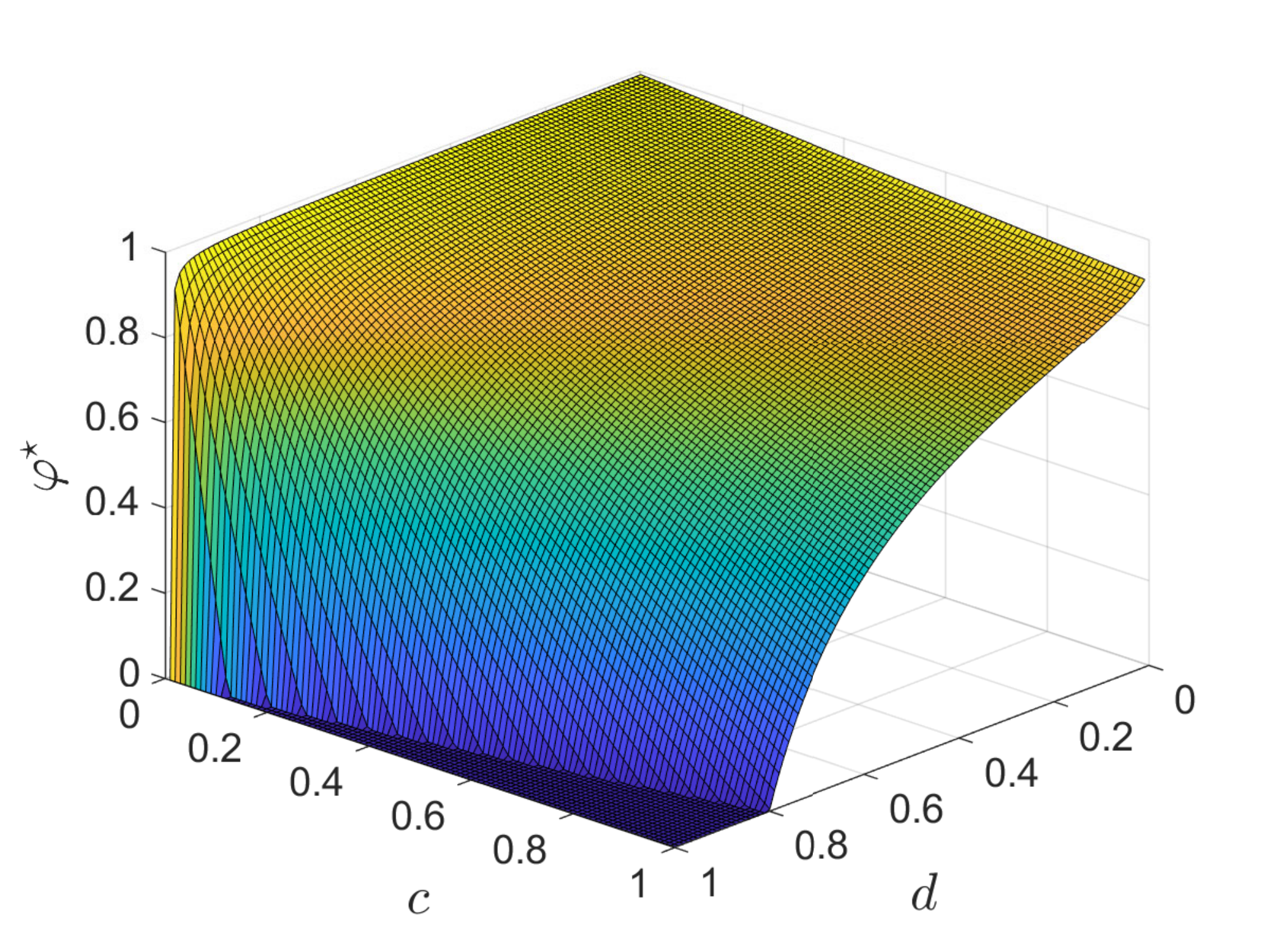}
	\caption{Optimal jamming probability for the  jammer without channel sensing $\varphi^\star$ as a function of  $c$ and $d$. Here, $X \sim \mathcal{N}(0,1)$.}
	\label{fig:agnosticjammerOptimalJammingProbabilityVSVar}
\end{figure}
{}

\begin{example} Consider $X \sim \mathcal{N}(0,1)$, and $c=d=1$. Since
\begin{equation}
2\int_1^\infty x^2f(x)\mathrm{d}x = 0.8012 < 1,
\end{equation}
the optimal jamming probability is
	$\varphi^\star = 0,$ 
which implies that $\gamma^\star$ and $\eta^\star$ are given by \cref{eq:saddle_pt_11,eq:saddle_pt_12}.
	Let $X \sim \mathcal{N}(0,2)$, and $c=d=1$. Since
\begin{equation}
2\int_1^\infty x^2f(x)\mathrm{d}x = 1.8378 > 1,
\end{equation}
the optimal jamming probability is $\varphi^\star = \tilde{\varphi} =0.7887$, which implies that $\gamma^\star$ and $\eta^\star$ are given by \cref{eq:saddle_pt_21,eq:saddle_pt_22}. \Cref{fig:agnosticjammerOptimalJammingProbabilityVSVar} shows the optimal jamming probability for the  jammer without channel sensing as a function of the communication and jamming costs, $c$ and $d$, for $X\sim \mathcal{N}(0,1)$\footnote{The code used to obtain the examples in this paper is available at GitHub (\url{https://github.com/mullervasconcelos/CDC22.git}).}.
\end{example}

\section{Reactive jammer}

When the jammer is able to sense if the channel is being used and uses this knowledge to adjust its probability of blocking the channel, its policy becomes:
\begin{equation} \label{jammingpolicy}
\varphi(0) \Equaldef \alpha 
\ \ \text{and}
 \ \ 
\varphi(1) \Equaldef \beta. 
\end{equation}
For brevity, let $\theta \Equaldef (\alpha,\beta)$.

\begin{figure}[t!]
	\centering
	\includegraphics[width=0.9\columnwidth]{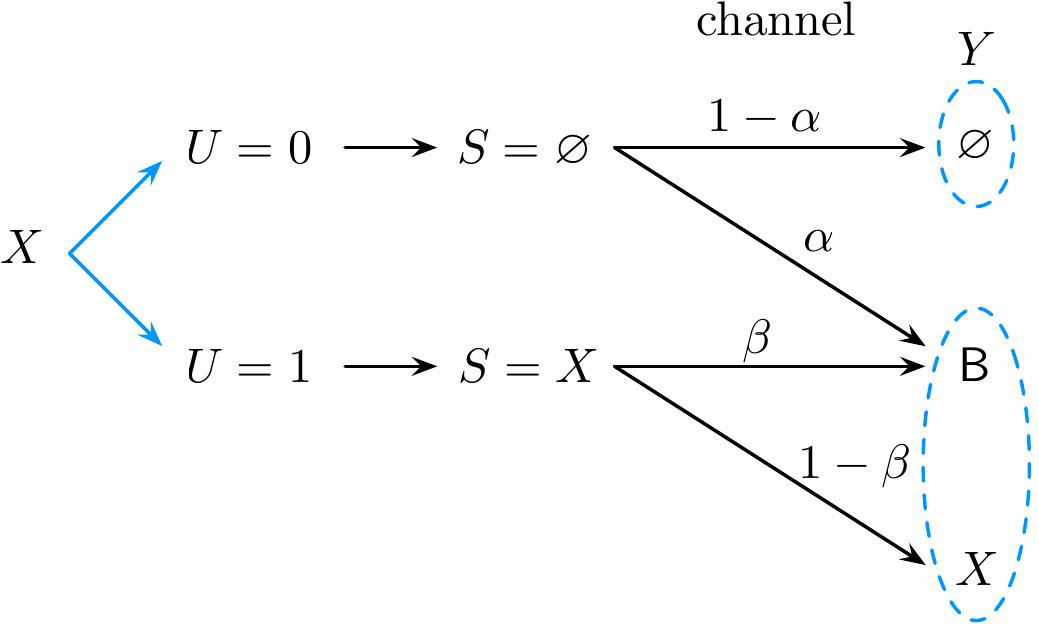}
	\caption{Signaling channel between the sensor and the receiver. The jammer controls the transition probabilities $\alpha$ and $\beta$. When $\alpha=0$, the channel is noiseless, i.e., the receiver can unequivocally decode whether $U=1$ or $U=0$ from the ouput signal $Y$.}
	\label{fig:channel}
\end{figure}

Notice that we allow the reactive jammer to block the channel even when the sensor is not transmitting. To the best of our knowledge, the existing literature on reactive jamming attacks precludes jamming when the channel is not being used. There is a reason why the jammer may engage in such counter-intuitive behavior: when the jammer only blocks a transmitted signal, it creates a noiseless binary (signaling) channel between the transmitter and the receiver, which may be exploited by the coordinator. If the jammer is allowed to ``block'' the channel when the user is not transmitting, such binary signaling channel is not noiseless anymore, because there will be uncertainty if the decision variable at the transmitter is zero or one. This scenario is illustrated in \cref{fig:channel}.

\vspace{5pt}


\begin{proposition}
\label{prop: best_threshold_reactive}
	For a fixed jamming policy $\varphi$ parametrized by $\theta\in [0,1]^2$, and a fixed estimation policy $\eta$ parametrized by $\hat{x}\in \mathbb{R}^2$, the optimal transmission policy is:
	\begin{multline}
		\gamma_{\eta,\varphi}^\star(x) = \mathbf{1}\Big( \beta (x-\hat{x}_1)^2  + c - d \beta \leq \\ \alpha (x-\hat{x}_1)^2  + (1-\alpha) (x-\hat{x}_0)^2 - d \alpha   \Big).
	\end{multline}
\end{proposition}

\vspace{5pt}

\begin{proof} 
For a reactive jammer, the random variables $X$ and $J$ are conditionally independent given $U$. Using the law of total expectation, and employing the estimation policy in \cref{eq: estimator}, the cost function can be reformulated as
\begin{multline}
	\mathcal{J}\big( (\gamma,\eta),\varphi \big) 
	=\mathbf{E}\big[(X-\hat{x}_0) \mid U=0 \big]\Pr(U=0) (1-\alpha) \\
	+ \mathbf{E}\big[(X-\hat{x}_1) \mid U=0 \big]\Pr(U=0) \alpha -d\Pr(U=0) \alpha \\
	+ \mathbf{E}\big[(X-\hat{x}_1) \mid U=1 \big]\Pr(U=1) \beta 
	+ (c -d\beta) \Pr(U=1),  
\end{multline}
which is equivalent to
\begin{multline} \label{eq:cost_reactive}
	\mathcal{J}\big( (\gamma,\eta),\varphi \big) = 
	\int_{\mathbb{R}} \big[\beta (x-\hat{x}_1)^2  + c - d \beta \big]\gamma(x)f(x)\mathrm{d} x \\
	+\int_\mathbb{R} \big[\alpha (x-\hat{x}_1)^2  + (1-\alpha) (x-\hat{x}_0)^2 - d \alpha\big]\big((1-\gamma(x)\big)f(x)\mathrm{d} x.
\end{multline}

For fixed $\hat{x} \in\mathbb{R}^2$ and $\theta\in[0,1]^2$, the transmission policy $\gamma$ that minimizes \cref{eq:cost_reactive} is obtained by comparing the arguments of the two integrals as follows:  $x \in \{\xi \mid \gamma^\star_{\eta,\varphi}(\xi) = 1\}$ if and only if 
\begin{equation}
	\beta (x-\hat{x}_1)^2  + c - d \beta \leq \\ \alpha (x-\hat{x}_1)^2  + (1-\alpha) (x-\hat{x}_0)^2 - d \alpha.
\end{equation}
\end{proof}

\vspace{5pt}

Given the optimal transmitter's strategy in \cref{prop: best_threshold_reactive}, the objective function becomes
\begin{multline}\label{eq:cost_reactive_new}{}
\mathcal{J}\big((\gamma_{\varphi,\eta}^{\star},\eta),\varphi\big) = \mathbf{E}\Big[ \min\big\{\beta(X-\hat{x}_1)^2 +c-d\beta , \\ \alpha(X-\hat{x}_1)^2 + (1-\alpha)(X-\hat{x}_0)^2 -d\alpha \big\} \Big] \Equaldef \tilde{\mathcal{J}}(\hat{x},\theta).
\end{multline}

Notice that for fixed $\hat{x}\in\mathbb{R}^2$, $\mathcal{J}$ is a concave function of $\theta$ for any pdf $f$. However, for fixed $\theta\in [0,1]^2$, $\mathcal{J}$ is non-convex in $\hat{x}$. Therefore, the game between the coordinator and the jammer reduces to the following minimax optimization problem:
\begin{equation}\label{eq:game}
\min_{\hat{x}\in \mathbb{R}^2} \max_{\theta \in [0,1]^2} \tilde{\mathcal{J}}(\hat{x},\theta),
\end{equation}
where $\tilde{\mathcal{J}}(\hat{x},\theta)$ is given by \cref{eq:cost_reactive_new}. 

Unfortunately, the structure of \cref{eq:cost_reactive_new} does not allow the same techniques we used to find a saddle-point equilibrium for the jammer without channel sensing. Instead, a useful alternative to the saddle-point (Nash-equilibrium) are the solutions that satisfy the first-order stationarity conditions of the minimization and the maximization problems, yielding in a larger class of policies, called First-order Nash-equilibria (FNE) \cite{Ostrovskii:2021,Nouiehed:2019,Facchinei:2003,Lin:2020}.


\vspace{5pt}

\begin{definition}[Approximate First-order Nash-equilibrium]
	Let $\varepsilon >0$. A pair of policies $(\hat{x}^{\star},\theta^{\star}) \in \mathbb{R}^2\times [0,1]^2$ is an approximate First-order Nash-equilibrium ($\varepsilon$-FNE) of the game if 
	\begin{equation} \label{eq:FNEcond}
		\|\nabla_{\hat{x}}\tilde{\mathcal{J}}(\hat{x}^{\star},\theta^{\star})\|_2 \le \varepsilon
	\end{equation}
	and
	\begin{equation}\label{eq:LP_condition}
		 \max_{\theta \in[0,1]^2}\langle \nabla_{\theta}\tilde{\mathcal{J}}(\hat{x}^{\star},\theta^{\star}),\theta-\theta^\star\rangle \leq \varepsilon.
	\end{equation}
\end{definition}	

\vspace{5pt}

\begin{proposition}
The function $\tilde{\mathcal{J}}(\hat{x},\theta)$ is  differentiable in $\hat{x}$ and $\theta$. Moreover, the partial gradients 
are 
\begin{multline} \label{eq:gradient_x_hat}
   \nabla_{\hat{x}}\tilde{\mathcal{J}}(\hat{x},\theta) 
    = \textbf{E}
    \Bigg[ 
     \begin{bmatrix}
        0 \\
        -2 \beta (X-\hat{x}_1)
    \end{bmatrix}\mathbf{1}(\gamma_{\eta,\varphi}^\star(X)=1)  \\
+
\begin{bmatrix}
    -2(1-\alpha) (X-\hat{x}_0)  \\
    -2 \alpha (X-\hat{x}_1) 
\end{bmatrix} \mathbf{1}(\gamma_{\eta,\varphi}^\star(X)=0)
\Bigg]
\end{multline}
and
\begin{multline} \label{eq:gradient_theta}
    \nabla_{\theta}\tilde{\mathcal{J}}(\hat{x},\theta) 
    =\textbf{E}\Bigg[ 
        \begin{bmatrix}
        0 \\
        (X-\hat{x}_1)^2 - d 
    \end{bmatrix}\cdot\mathbf{1}(\gamma_{\eta,\varphi}^\star(X)=1) \\
+
    \begin{bmatrix}
        (X-\hat{x}_1)^2  - (X-\hat{x}_0)^2 - d   \\
        0
        \end{bmatrix} \mathbf{1}(\gamma_{\eta,\varphi}^\star(X)=0)\Bigg].
\end{multline} 
\end{proposition}

\vspace{5pt}

\begin{proof}
This result follows from the Leibniz rule and is omitted due to space constraints.
\end{proof}

\vspace{5pt}

\subsection{Optimization algorithm}

To obtain a pair of $\varepsilon$-FNE to the game in \cref{eq:game}, we alternate between a \textit{projected gradient ascent} (PGA) step for the inner optimization problem; and a \textit{convex-concave procedure} (CCP) step for the outer optimization problem.

We start with the description of the PGA step at a point $(\hat{x}^{(k)},\theta^{(k)})$:
\begin{equation}
	\theta^{(k+1)}= \mathcal{P}_{[0,1]^2}\big(\theta^{(k)} + \lambda_k\, \nabla_{\theta}\tilde{\mathcal{J}}(\hat{x}^{(k)},\theta^{(k)}\big),
\end{equation}
where $\{\lambda_k\}$ is a step-size sequence (e.g. $\lambda_k=0.1/\sqrt{k}$) and the projection operator is defined as  $\mathcal{P}_{[0,1]^2}(\theta):=\min_{\bar{\theta} \in [0,1]^2} \| \bar{\theta}-\theta \|_2$, which is equal to
\begin{equation}
	\mathcal{P}_{[0,1]^2} \bigg( 
	\begin{bmatrix}
		\alpha \\
		\beta
	\end{bmatrix} \bigg) = \bigg[ 
\begin{array}{c}
\max\big\{0, \min\{1, \alpha\} \big\} \\
\max\big\{0, \min\{1, \beta\} \big\}
\end{array} \bigg].
\end{equation}


To update $\hat{x}^{(k)}$ for a fixed $\theta^{(k+1)}$, we use the property that \cref{eq:cost_reactive_new} can be decomposed as a difference of convex functions (DC decomposition). Using the  
DC decomposition we obtain a specialized descent algorithm  \cite{Yuille:2003}, which is guaranteed to converge to stationary points of \cref{eq:cost_reactive_new} for a fixed $\theta^{(k+1)}$\cite{Lipp:2016}.
\textcolor{black}{CCP uses more information about the structure of the objective function than standard Gradient Descent methods, often leading to faster convergence \cite{Yuille:2003}.}   

\begin{algorithm}[t]
	\caption{PGA-CCP algorithm}
	\label{alg: I}
	\begin{algorithmic}[1]
		\REQUIRE  PDF $f$, transmission cost $c$, jamming cost $d$   
		\ENSURE   Estimated result $\hat{x}^{\star}$ and $\theta^{\star}$
		\STATE
		Initialize 
		$k\gets0,$ $\varepsilon,$ $\hat{x}^{(0)}$ and $\theta^{(0)}$
		\REPEAT
		\STATE $\theta^{(k+1)}=   \mathcal{P}_{[0,1]^2}\big(\theta^{(k)} + \lambda_k\,\nabla_{\theta}\tilde{\mathcal{J}}(\hat{x}^{(k)},\theta^{(k)}\big)$
		\STATE $\hat{x}^{(k+1)} = \mathcal{A}^\dagger(\theta^{(k+1)}) \, g(\hat{x}^{(k)},\theta^{(k+1)})$
		\STATE $k \gets k+1$
		\UNTIL $\varepsilon$-FNE conditions (\cref{eq:FNEcond,eq:LP_condition}) are satisfied 
	\end{algorithmic}
\end{algorithm}

Notice that:
\begin{equation}
\tilde{\mathcal{J}}(\hat{x},\theta)= \mathcal{F}(\hat{x},\theta)-\mathcal{G}(\hat{x},\theta),
\end{equation}
where
\begin{multline}
	\mathcal{F}(\hat{x},\theta) \Equaldef (1-\alpha) \hat{x}_0^2 +(\alpha+\beta) \hat{x}_1^2 +(1+\beta)\sigma_X^2+c-d(\alpha+\beta),
\end{multline}
and
\begin{multline}
	\mathcal{G}(\hat{x},\theta) \Equaldef \mathbf{E}\Big[ \max\big\{\beta(X-\hat{x}_1)^2 +c-d\beta , \\ \alpha(X-\hat{x}_1)^2 + (1-\alpha)(X-\hat{x}_0)^2 -d\alpha \big\} \Big].
\end{multline}

The CCP for computing a local minima for the outer optimization problem is given by
\begin{equation} \label{eq: x_optimization_problem}
	\hat{x}^{(k+1)}= \arg \min_{\hat{x}} \left\{ \mathcal{F}(\hat{x},\theta^{(k+1)})- 	\mathcal{G}_a(\hat{x},\theta^{(k+1)};\hat{x}^{(k)}) \right\},
\end{equation}
where $\mathcal{G}_a(\hat{x},\theta^{(k+1)};\hat{x}^{(k)})$ is the affine approximation of $\mathcal{G}(\hat{x},\theta^{(k+1)})$ with respect to $\hat{x}$ at $\hat{x}^{(k)}$, while keeping $\theta^{(k+1)}$ fixed, i.e.,
\begin{multline}
	\mathcal{G}_a(\hat{x},\theta^{(k+1)};\hat{x}^{(k)})= \mathcal{G}(\hat{x}^{(k)},\theta^{(k+1)}) \\+ g^{T}(\hat{x}^{(k)},\theta^{(k+1)}) (\hat{x}-\hat{x}^{(k)})
\end{multline}
and $g(\hat{x},\theta)$ is the gradient of $\mathcal{G}(\hat{x},\theta)$  with respect to $\hat{x}$.

Because $\mathcal{F}$ is a quadratic function of $\hat{x}$ for a fixed $\theta$, we may use the first-order necessary optimality condition of problem \cref{eq: x_optimization_problem} to find the recursion for $\hat{x}^{(k+1)}$ in closed form:
\begin{equation}
	\nabla_{\hat{x}} \mathcal{F}(\hat{x}^{(k+1)},\theta^{(k+1)})=g(\hat{x}^{(k)},\theta^{(k+1)}).
\end{equation}
The partial gradient of $\mathcal{F}(\hat{x},\theta)$ with respect to $\hat{x}$ is
\begin{equation}
	\nabla_{\hat{x}} \mathcal{F}(\hat{x},\theta)=\left[ 
	\begin{array}{c}
		2(1-\alpha) \hat{x}_0 \\
		2(\alpha+\beta) \hat{x}_1
	\end{array} \right].
\end{equation}
 
 The partial gradient of $\mathcal{G}(\hat{x},\theta)$  with respect to $\hat{x}$ is
 \begin{multline}
 	g(\hat{x},\theta) 
    = \textbf{E}
    \Bigg[ 
     \begin{bmatrix}
        0 \\
        -2 \beta (X-\hat{x}_1)
    \end{bmatrix}\mathbf{1}(\gamma_{\eta,\varphi}^\star(X)=0)  \\
+
\begin{bmatrix}
    -2(1-\alpha) (X-\hat{x}_0)  \\
    -2 \alpha (X-\hat{x}_1) 
\end{bmatrix} \mathbf{1}(\gamma_{\eta,\varphi}^\star(X)=1)
\Bigg].
 \end{multline}
 
Finally, define $\mathcal{A}:[0,1]^2\rightarrow \mathbb{R}^{2\times 2}$ as
 \begin{equation}
 	\mathcal{A}(\theta)= \left[ 	\begin{array}{cc}
 		2(1-\alpha) & 0\\
 		0 & 2(\alpha+\beta)
 	\end{array} \right],
 \end{equation}
and $\mathcal{A}^\dagger$  denotes its Moore-Penrose pseudo-inverse. Then, the update of CCP can be compactly represented as
 \begin{equation}
 	\hat{x}^{(k+1)} = \mathcal{A}^\dagger\big{(}\theta^{(k+1)}\big{)} \, g\big{(}\hat{x}^{(k)},\theta^{(k+1)}\big{)}.
 \end{equation}

 The PGA-CCP algorithm is presented in \cref{alg: I} and its empirical efficacy is demonstrated in the following example.



\begin{example}
Let $X \sim \mathcal{N}(0,\sigma^2)$, and $c=d=1$. We set $\varepsilon = 10^{-5}$. Our algorithm results in the pairs of $\varepsilon$-FNE in table \ref{table: FNE pairs}\,\footnote{The code we used to implement the PGA-CCP algorithm is available at GitHub (\url{https://github.com/mullervasconcelos/CDC22.git}).}.
\begin{table} [!t]
\caption{$\varepsilon$-FNE obtained using the PGA-CCP algorithm}
\begin{center}
\begin{tabular}{c c c c c}  
 \hline
 $\sigma^2$ & $\alpha^\star$ & $\beta^\star$ & $\hat{x}_0^\star$ & $\hat{x}_1^\star$ \\
 \hline\hline
 $1$ & $0.0760$  &  $0.3172$ &   $0.5169$ &   $-0.4831$ \\ 
 $2$ & $0.0350$  &  $0.1572$ &   $0.7030$  &  $-0.4338$ \\
 $3$ & $0.0136$  &  $0.0937$ &   $0.7618$ &  $-0.4204$ \\
 $4$ & $0.0040$ &   $0.0634$ &  $0.7894$ &   $-0.4148$ \\
 $5$ & $0$   & $0.0475$ &  $0.8082$ &   $-0.4039$ \\
 \hline
\end{tabular} \label{table: FNE pairs}
\end{center}
\end{table}
\textcolor{black}{Figure \ref{fig:example_detail} shows the optimal jamming probabilities $\alpha^\star$ and $\beta^\star$ as a function of $\sigma^2$.} Notice that the probability that the jammer will block the channel even when the sensor does not transmit is nonzero.
\textcolor{black}{Figure \ref{fig:OptimalThresholding} shows that the optimal transmission policy is asymmetric for $c=1,d=1$, and $X \sim \mathcal{N}(0,1)$. Figure \ref{fig:comparison} shows the convergence to an $\varepsilon$-FNE for $c=1,d=1$, and $X \sim \mathcal{N}(0,1)$ using the PGA-CCP (this paper) and the Gradient Descent Ascent (GDA) \cite{Lin:2020} algorithms. \textcolor{black}{The step size for PGA-CCP is set to be $\lambda=0.1$ and the step sizes for GA and GD in GDA are set to be $\lambda_{\mathrm{GA}}=0.1$ and $\lambda_{\mathrm{GD}}=0.01$, respectively\footnote{\textcolor{black}{For the sake of fairness, the step sizes for the GA update of both algorithms are set to be the same. Due to the asymmetric nature of nonconvex-concave problems, GDA requires a time-scale separation in the step-sizes to avoid convergence to limit cycles or even divergence \cite{Lin:2020}. The reason for $\lambda_{\mathrm{GA}}>\lambda_{\mathrm{GD}}$ is that the inner maximization problem has better structure (the cost function is concave) \cite{Lin:2020}.}}.}  Our numerical results show that PGA-CCP converges with a rate approximately $6$ times faster than the algorithm in \cite{Lin:2020}.}

\begin{figure}[t]
	\centering
	\includegraphics[width=0.9\columnwidth]{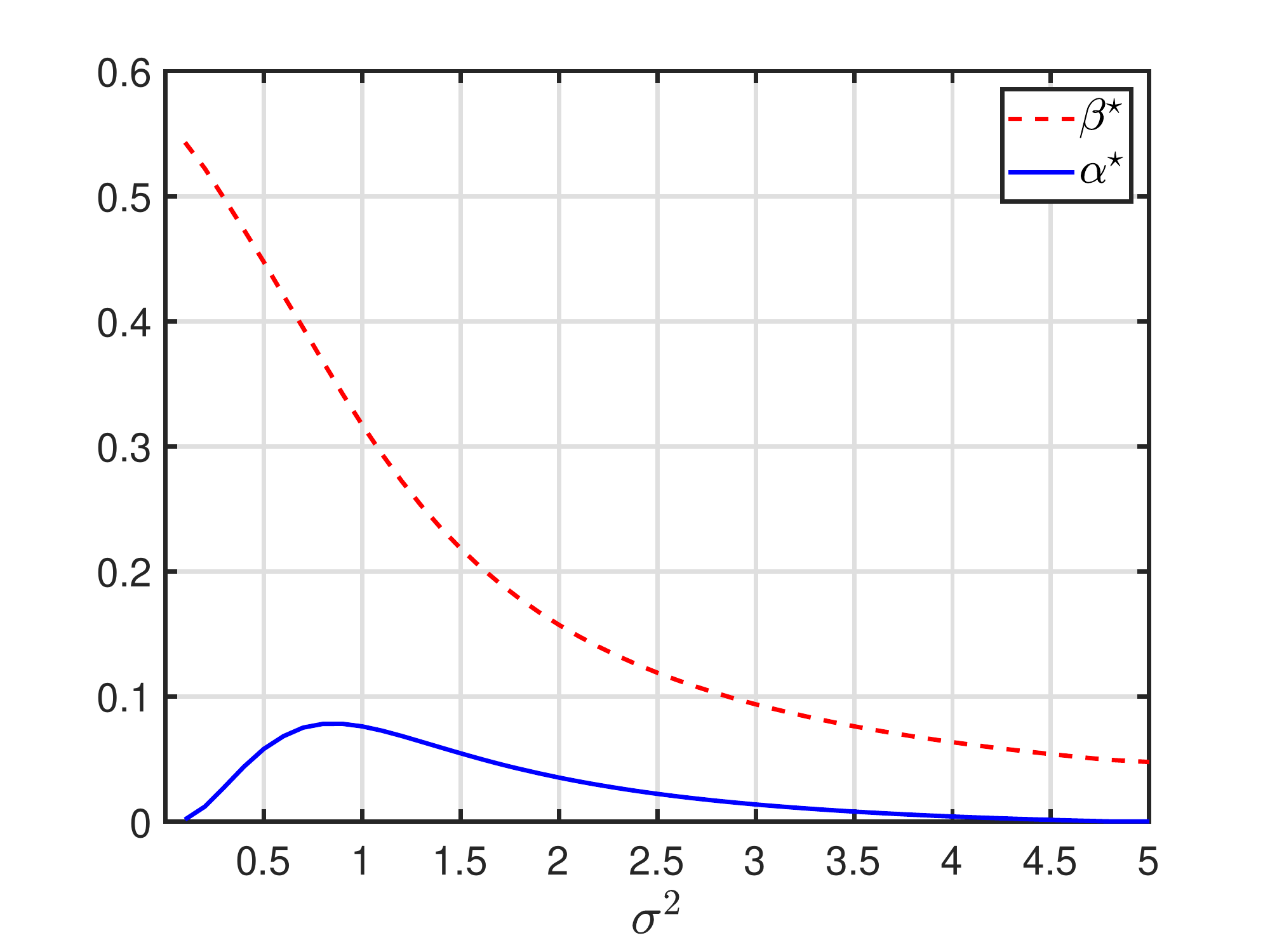}
	\caption{Optimal jamming probabilities $\alpha^\star$ and $\beta^\star$ as a function of $\sigma^2$. Here, $c=1,d=1$, and $X \sim \mathcal{N}(0,\sigma^2)$.}
	\label{fig:example_detail}
\end{figure}

\begin{figure}[t]
	\centering
	\includegraphics[width=0.95\columnwidth]{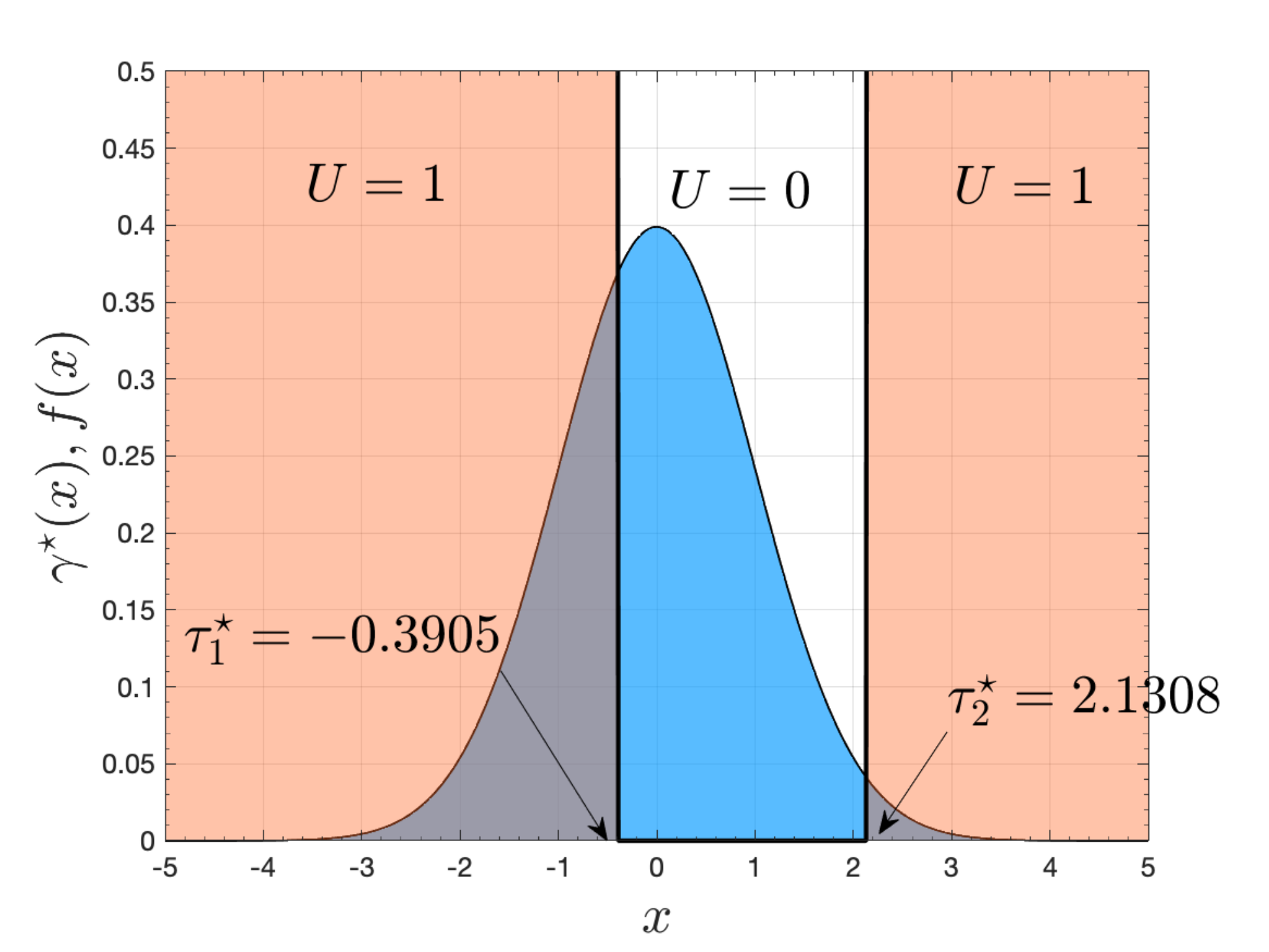}
	\caption{Optimal transmission policy for $c=1,d=1$, and $X \sim \mathcal{N}(0,1)$.}
	\label{fig:OptimalThresholding}
\end{figure}

\begin{figure}[t]
    \centering
    \includegraphics[width=0.95\columnwidth]{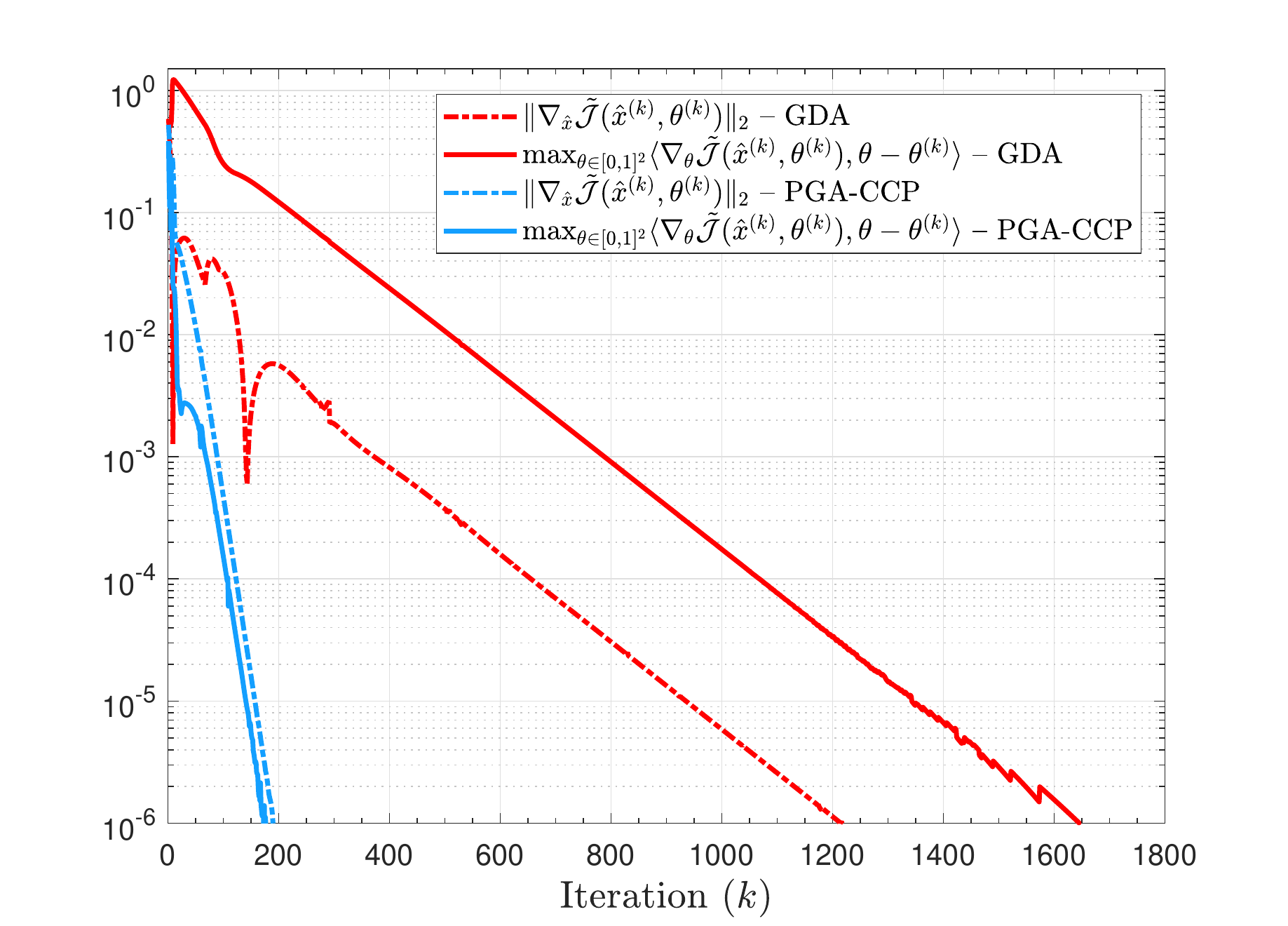}
    \caption{Convergence of PGA-CCP vs. GDA for $c=1,d=1$, and $X \sim \mathcal{N}(0,1)$.}
    \label{fig:comparison}
\end{figure}

\end{example}

%
%
%


\section{Conclusions and future work}
We have studied a zero-sum signaling game with asymmetric information involving a sensor, a jammer and an estimator. Two scenarios were considered: jamming with and without channel sensing. For the jammer without channel sensing, we have shown that under a symmetry and unimodality assumption of the observation's pdf, there exists a saddle-point equilibrium, where the optimal transmission policy at the sensor is of the symmetric threshold type. For a reactive jammer, the objective function does not admit a similar result. Instead, we exploit the structure of outer and inner-optimization problem to obtain an algorithm to find $\varepsilon$-FNE. 
There are many interesting research directions for future work. The first possible extension is the case when there are $n$ sensors sharing the network over a channel that can support $k<n$ packets. Additionally, it is important to prove the convergence of the PGA-CCP algorithm and the characterization of its convergence rate. Finally, study how the coordinator and the jammer learn to play in equilibrium if they do not have access to each other's costs.


\bibliographystyle{IEEEtran}

\bibliography{IEEEabrv,refs}

%
%
%


\end{document}